\documentclass[11pt]{article}
\def \d {{\rm d}}

\begin{document}

\title{The characteristic initial value problem for colliding
plane waves: The linear case}

\author{J. B. Griffiths\thanks{E--mail: {\tt J.B.Griffiths@Lboro.ac.uk}} \ and M.
Santano-Roco 
\\ \\ 
Department of Mathematical Sciences, Loughborough University, \\ 
Loughborough, Leics. LE11 3TU, U.K. \\ }

\date{\today}
\maketitle

\begin{abstract}
\noindent
The physical situation of the collision and subsequent interaction of plane
gravitational waves in a Minkowski background gives rise to a well-posed
characteristic initial value problem in which initial data are specified on
the two null characteristics that define the wavefronts. In this paper, we
analyse how the Abel transform method can be used in practice to solve this
problem for the linear case in which the polarization of the two gravitational
waves is constant and aligned. We show how the method works for some known
solutions, where problems arise in other cases, and how the problem can always
be solved in terms of an infinite series if the spectral functions for the
initial data can be evaluated explicitly. 
\end{abstract}

\section{Introduction}
Many classes of explicit exact solutions are now known which model the collision
and subsequent interaction between plane gravitational waves with distinct
wavefronts which propagate into a Minkowski background (for a review see
\cite{Griff91}). These solutions have invariably been obtained using an indirect
approach which starts by finding a family of solutions of the field equations in
the interaction region. The arbitrary parameters characterising the family are
then restricted to specific ranges in order to satisfy the appropriate junction
conditions for colliding plane waves. Finally, the restricted class of solutions
obtained is extended backwards to determine the initial approaching waves that
give rise to that particular family of solutions. This approach has enabled the
general structure of this class of solutions to be understood. However, little
progress has been made using the direct approach in which the approaching plane
waves are specified initially and the subsequent interaction between them
following their collision is then determined by solving the associated
characteristic initial value problem.

It is known \cite{Szek72}--\cite{Yurts88} that, in the vacuum (linear) case in
which the approaching waves have constant and aligned polarization, the
characteristic initial value problem may be solved theoretically using Riemann's
method. Unfortunately, this method involves integrals which cannot be evaluated
explicitly for appropriate initial data. As an alternative, Hauser and Ernst
\cite{HauErn89a} have formulated a different method for solving the initial value
problem in the linear case by making use of Abel transforms. Unfortunately,
neither of these methods can be obviously generalised to the nonlinear case.
However, some progress in treating this general case has recently been achieved.
Hauser and Ernst \cite{HauErn89b}--\cite{HauErn91} reformulated the corresponding
nonlinear (vacuum) problem as a homogeneous Hilbert problem in a complex plane.
Using this, they have proved that a solution of this problem exists, but their
method has still not been used in practice to determine an explicit solution.
However, another approach to boundary or initial value problems, extended to
include the electromagnetic case, had previously been formulated \cite{Aleks85},
\cite{Aleks87} in terms of the monodromy transform method. Unfortunately,
problems involving singularities at the point of collision are encountered when
applying this method to the case of colliding plane waves. Nevertheless, these
problems can be overcome. By extending the method to include dynamical monodromy
data, a new method for solving the initial value problem has been achieved. This
has been outlined in \cite{AleGri01} and is described in detail in
\cite{AleGri02}.

In the present work we concentrate on the linear case in which the polarization
of the two initial gravitational waves is constant and aligned. We formulate the
characteristic initial value problem for this situation in detail. We then
demonstrate how the Abel transform method can be implemented in practice by
deriving some well known solutions. Of course, to implement the Abel transform
method to determine an explicit exact solution in the interaction region for
given initial data, it is necessary to evaluate a number of integrals. We
describe the particular difficulties that arise in attempting to implement this
method for some physically interesting initial data. Finally, we demonstrate how
the exact solution in the interaction region corresponding to arbitrary initial
data can be constructed in terms of an infinite series involving hypergeometric
functions.

\section{A collision of gravitational plane waves}

First, we recall that a plane gravitational wave with linear polarization and
arbitrary profile $h(u)$ can be described by a metric in the Brinkmann form 
 \begin{equation}
 \d s^2 =2\,\d u\,\d r -\d X^2 -\d Y^2
+h(u)(X^2-Y^2)\,\d u^2. 
 \label{Brinkmann}
 \end{equation} 
 The wave surfaces for this wave are given by \ $u=$~const. \ It may be noted that
the profile function appears explicitly in the metric. This has a distinct
disadvantage when the profile takes a distributional form as, for example, in a
description of impulsive waves. For this and a variety of other reasons, it will
be convenient to transform the metric (\ref{Brinkmann}) to the Rosen form 
 \begin{equation}
 \d s^2 =2\,\d u\,\d v -P^2\,\d x^2 -Q^2\,\d y^2, 
 \label{Rosen}
 \end{equation} 
 which is always continuous, and involves two null coordinates $u$ and $v$. This
can be achieved using  the transformation \ $X=P(u)x$, \ $Y=Q(u)y$, \
$r=v+{1\over2}PP'x^2+{1\over2}QQ'y^2$, \ where 
 \begin{equation}
P''+h(u)P=0 \qquad {\rm and} \qquad Q''-h(u)Q=0, 
 \label{PQdef}
 \end{equation}
 so that the only field equation is 
 \begin{equation}
 {P''\over P} +{Q''\over Q}=0. 
 \label{Rosenequation}
 \end{equation}
 Using a natural tetrad, the only non-zero
component of the Weyl tensor for the metric (\ref{Rosen}) is given by 
 \begin{equation}
 \Psi_4={1\over2}\left( {Q''\over Q}-{P''\over P} \right) =h(u). 
 \end{equation}

Now consider a gravitational wave (with constant polarization) which has a
distinct wavefront. We can label the wavefront by $u=0$, so that $h(u)=0$ and
the space is Minkowski for $u<0$. The wave may then be considered to have
arbitrary profile $h(u)$ for $u\ge0$, and be represented by the line element
(\ref{Rosen}). We can also consider a second gravitational wave of a similar type
to propagate into the initial Minkowski space from the opposite spatial direction.
This may be considered to have a wavefront on the null hypersurface $v=0$ using
the same notation as in (\ref{Rosen}). If this wave also has constant
polarization (with the polarizations of the two waves aligned), the region $u<0$,
$v\ge0$ can also be described by a metric of the form (\ref{Rosen}) but with $P$
and $Q$ being functions of $v$ instead of $u$.

In this way, we have set up initial conditions for the collision of two plane
waves with constant aligned polarization and arbitrary profiles following the
wavefronts. The space-time contains four distinct regions --- a background
Minkowski region~I ($u<0$, $v<0$) which possesses the metric 
 \begin{equation}
 \d s^2=2\d u\,\d v-\d x^2-\d y^2, 
 \label{Minkowski}
 \end{equation} 
 two wave regions II and III ($u\ge0$, $v<0$ and $u<0$, $v\ge0$ respectively),
and a region~IV ($u\ge0$, $v\ge0$) which is subsequent to the collision of the
two waves and in which they interact. The general structure of such space-times
is well known and illustrated in figure~1.

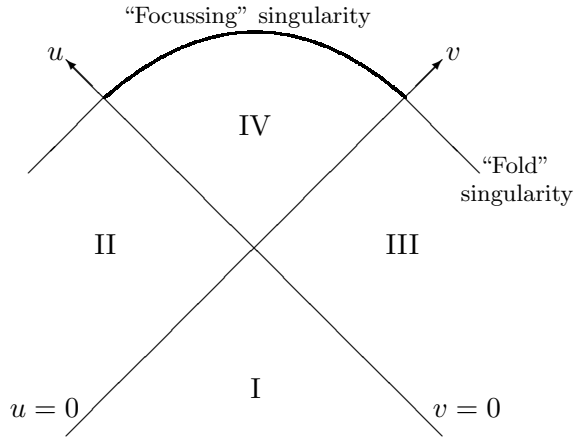
\begin{figure}[hpt]      
\begin{center}
\setlength{\unitlength}{0.25mm}
\begin{picture}(200,260)(-100,-100)
\put(-100,-100){\vector(1,1){200}}
\put(100,-100){\vector(-1,1){200}}
\put(-80,80){\line(-1,-1){40}}
\put(80,80){\line(1,-1){40}}
\put(-110,100){$u$}
\put(102,100){$v$}
\put(-130,-90){$u=0$}
\put(95,-90){$v=0$}
\put(-2,-80){I}
\put(-85,-3){II}
\put(70,-3){III}
\put(-8,60){IV}
\put(-70,120){\fontsize{9}{14}\selectfont ``Focussing'' singularity}
\put(120,40){\fontsize{9}{14}\selectfont ``Fold''}
\put(110,25){\fontsize{9}{14}\selectfont singularity}
\thicklines
\qbezier(-80,80)(0,150)(80,80)
\end{picture}
\end{center}

\caption{ The structure of colliding plane wave space-times. Region~I is the
background Minkowski space, regions II and~III contain the approaching plane
waves, and region~IV represents the interaction region following the collision.
The singularity structure following the collision is known from the study of
families of exact solutions and is described in \cite{Griff91}.}
\end{figure}

For convenience, we will denote functions and parameters associated with the
region~II with the subscript~$+$, and functions and parameters associated with the
region~III with the subscript~$-$. Thus, the two waves which propagate into the
background have profile functions \ $\Psi_4=h_{\scriptscriptstyle+}(u)$ \ for \
$u\ge0$, $v<0$, \ and \ $\Psi_0=h_{\scriptscriptstyle-}(v)$ \ for \ $v\ge0$,
$u<0$.

The problem we now address is to initially specify the initial profile functions
$h_{\scriptscriptstyle+}(u)$ and $h_{\scriptscriptstyle-}(v)$ and then to attempt
to determine the metric to which this gives rise in the interaction region~IV.
This is a well posed characteristic initial value problem for which a unique
solution is known to exist. However, as will be seen, it is not a simple matter to
construct the solution explicitly.

\section{Analysis of the initial data}

It is found to be convenient to continue to use two future-pointing null
coordinates $u$ and $v$ throughout the space-time, with the four distinct regions
being separated by two null hypersurfaces \ $u=0$ \ and \ $v=0$ \ which represent
the wavefronts of the approaching waves. 
The background region~I ($u<0$, $v<0$) ahead of each wave is taken to be part of
Minkowski space with the metric~(\ref{Minkowski}).

{\bf Region~II} ($u\ge0$, $v<0$) contains one of the approaching plane waves. It
is found to be convenient to represent this by the line element 
 \begin{equation}
 \d s^2=2\d u\,\d v -e^{-U_{\scriptscriptstyle+}} 
\big(e^{V_{\scriptscriptstyle+}}\,\d x^2
+e^{-V_{\scriptscriptstyle+}}\,\d y^2\big) 
 \label{planewave}
 \end{equation}
 in which the metric functions \ $U_{\scriptscriptstyle+}
=-\log{P_{\scriptscriptstyle+}}-\log{Q_{\scriptscriptstyle+}}$, \ and \
$V_{\scriptscriptstyle+} =\log{P_{\scriptscriptstyle+}}
-\log{Q_{\scriptscriptstyle+}}$ \ depend on $u$ only. For any given initial
profile function $h_{\scriptscriptstyle+}(u)$, they are constructed from the
corresponding solutions of (\ref{PQdef}), and must satisfy the single equation 
 \begin{equation}
 2U_{{\scriptscriptstyle+},uu} ={U_{{\scriptscriptstyle+},u}}^2
+{V_{{\scriptscriptstyle+},u}}^2, 
 \label{condition1}
 \end{equation}
 which corresponds to (\ref{Rosenequation}). The amplitude of the initial
gravitational wave is given by 
 \begin{equation}
 h_{\scriptscriptstyle+}(u)=-{\textstyle{1\over2}}
\big(V_{{\scriptscriptstyle+},uu}
-U_{{\scriptscriptstyle+},u}V_{{\scriptscriptstyle+},u}\big) 
 \label{h(u)1}
 \end{equation}

In order to join this region smoothly with the metric (\ref{Minkowski}) across the
null hypersurface \ $u=0$, \ without introducing impulsive components in the Ricci
tensor, it is necessary that \ $U_{\scriptscriptstyle+}(0)=0$, \
$U_{\scriptscriptstyle+,u}(0)=0$, \ and \ $V_{\scriptscriptstyle+}(0)=0$. \ It is
then clear from (\ref{condition1}) that $U_{\scriptscriptstyle+}$ and
$U_{{\scriptscriptstyle+},u}$ must be monotonically increasing functions. Hence
$e^{-U_{\scriptscriptstyle+}}$ is a monotonically decreasing function in this
region and can be expanded in the form 
 \begin{equation}
 e^{-U_{\scriptscriptstyle+}}
=1-(c_{\scriptscriptstyle+}u)^{n_{\scriptscriptstyle+}}
+o(u^{n_{\scriptscriptstyle+}}) 
 \end{equation}
 near the wavefront \ $u=0$, \ where \ $c_{\scriptscriptstyle+}>0$ \ and
$n_{\scriptscriptstyle+}\ge2$ are constants. It then follows from
(\ref{condition1}) that, near the wavefront, $V_{\scriptscriptstyle+}$ must be
such that 
 \begin{equation}
 {V_{{\scriptscriptstyle+},u}}^2 =
2n_{\scriptscriptstyle+}(n_{\scriptscriptstyle+}-1) 
c_{\scriptscriptstyle+}^{n_{\scriptscriptstyle+}} u^{(n_+-2)} 
+o(u^{(n_+-2)}), 
 \label{ucondition}
 \end{equation} 
 and the amplitude function behaves as 
 \begin{equation}
 h_{\scriptscriptstyle+}(u)= \pm {\textstyle{1\over4}}
\sqrt{2n_{\scriptscriptstyle+} (n_{\scriptscriptstyle+}-1)}\,
(n_{\scriptscriptstyle+}-2)\,
c_{\scriptscriptstyle+}^{n_{\scriptscriptstyle+}/2}\, u^{(n_+/2)-2}
+o(u^{(n_+/2)-2}), 
 \end{equation} 
 where the sign is opposite to that of $V_{{\scriptscriptstyle+},u}$. (It is
assumed that all functions are sufficiently smooth as $u\to0$.) It follows from
this that the parameters $n_{\scriptscriptstyle+}$ and $c_{\scriptscriptstyle+}$
are determined uniquely from the initial data --- specifically from the behaviour
of
$h_{\scriptscriptstyle+}(u)$ near the wavefront. (The case when \
$n_{\scriptscriptstyle+}=2$ \ generally contains an impulsive component on the
wavefront $u=0$, and needs to be treated a little more carefully.)

Now, it is always possible to relabel the null hypersurfaces using an arbitrary
$C^2$ transformation \ $u\to\tilde u=\tilde u(u)$. \ In practice, this freedom is
almost always required in the construction of explicit solutions. However, it has
the effect of replacing the term \ $2\,\d u\,\d v$ \ in the line element
(\ref{planewave}) by \ $2e^{-M_{\scriptscriptstyle+}(u)}\d u\,\d v$. \ Further,
in order to avoid introducing a coordinate singularity and to maintain continuity
with the background metric in the form (\ref{Minkowski}), it is necessary that \
$\d u/\d\tilde u=1$ \ on the wavefront labelled $u=0$,
$\tilde u=0$. Using this freedom, it is possible to choose some particular form
for $e^{-U_{\scriptscriptstyle+}}$, such as 
 \begin{equation}
 e^{-U_{\scriptscriptstyle+}}
=1-(c_{\scriptscriptstyle+}u)^{n_{\scriptscriptstyle+}}. 
 \label{gauge1}
 \end{equation} 
 It is significant that the power of the leading term in the expansion of the
metric function $U_{\scriptscriptstyle+}$ cannot be changed. i.e.
$n_{\scriptscriptstyle+}$ is an essential parameter which is determined by the
initial data. Physically, this parameter determines the character of the
wavefront; mathematically, it determines the degree of continuity of
$U_{\scriptscriptstyle+}$ on the wavefront \ $u=0$. \

Retaining complete generality at this stage it is convenient to put 
 \begin{equation}
 e^{-U_{\scriptscriptstyle+}}={\textstyle{1\over2}}+f(u), 
 \label{Uplus}
 \end{equation}
 and then to use $f$ as a coordinate in place of $u$. It may be observed that
this function is given by \ $f=P_{\scriptscriptstyle+} Q_{\scriptscriptstyle+}
-{\textstyle{1\over2}}$, \ which is explicitly determined by the initial data. To
maintain the character of the wavefront at \ $u=0$, \ it is necessary that $f(u)$
behaves as 
 $$ f(u)={\textstyle{1\over2}}
-(c_{\scriptscriptstyle+}u)^{n_{\scriptscriptstyle+}}
+o(u^{n_{\scriptscriptstyle+}}) $$ 
 near the wavefront. In fact it is always possible (although not always
convenient) to use the above coordinate freedom to introduce a special gauge
(\ref{gauge1}) such that 
 \begin{equation}
 f(u)={\textstyle{1\over2}}
-(c_{\scriptscriptstyle+}u)^{n_{\scriptscriptstyle+}}. 
 \label{f}
 \end{equation}
 The additional metric function
$M_{\scriptscriptstyle+}$, which is normally introduced with any special choice of
gauge, is then determined by the equation 
 \begin{equation}
 M_{{\scriptscriptstyle+}f}=-{f''\over{f'}^2} +{1\over2(f+{1\over2})}
-{\textstyle{1\over2}} (f+{\textstyle{1\over2}})
{V_{{\scriptscriptstyle+}f}}^2, 
 \label{Mf}
 \end{equation}
 which replaces (\ref{condition1}). This equation is critical in the colliding
plane wave problem as it contains a singular term on the wavefront on which \
$f'=0$, and this must be counterbalanced by an appropriate singular behaviour in
the function $V_{\scriptscriptstyle+}$. In order for $M_{\scriptscriptstyle+}$ to
be continuous across the boundary \ $u=0$, \ on which \ $f'=0$, \ it is necessary
that $V_{\scriptscriptstyle+}$ satisfies (\ref{ucondition}). Rewritten in terms
of the function $f$, this condition can be expressed as 
 \begin{equation}
 \lim_{\scriptstyle f\to1/2}
\left[({\textstyle{1\over2}}-f) (V_{{\scriptscriptstyle+}f})^2
\right] =2k_{\scriptscriptstyle+},
 \label{shockcondition}
 \end{equation}
 where \ $k_{\scriptscriptstyle+}=1-1/n_{\scriptscriptstyle+}$ \ i.e. 
 $$ {\textstyle{1\over2}}\le k_{\scriptscriptstyle+}<1. $$

\medskip\goodbreak
{\bf Region~III} ($u<0$, $v\ge0$), contains the other approaching plane
gravitational wave with the same (i.e.~aligned) polarization. The metric can be
taken in the same form as that for region~II but with the roles of $u$ and $v$
reversed. It is here convenient to put 
 \begin{equation}
 e^{-U_{\scriptscriptstyle-}}={\textstyle{1\over2}}+g(v), 
 \label{Vplus}
 \end{equation}
 and to use $g$ as a coordinate in place of $v$. Again, it is sometimes
convenient to adopt a gauge such that 
 \begin{equation}
 g(v)={\textstyle{1\over2}}
-(c_{\scriptscriptstyle-}v)^{n_{\scriptscriptstyle-}}. 
 \label{g}
 \end{equation}
 It may be noted that the amplitude of the two waves and the character of the
wavefronts are determined by the constants $c_{\scriptscriptstyle+}$, 
$c_{\scriptscriptstyle-}$, $n_{\scriptscriptstyle+}$ and
$n_{\scriptscriptstyle-}$. The latter two are fixed by the initial data, but it
is always possible to make a boost \ $u\to c\tilde u$, $v\to c^{-1}\tilde v$ \
such that either $c_{\scriptscriptstyle+}$ or $c_{\scriptscriptstyle-}$ can be
made equal to unity. (Alternatively, they could be made equal.)

 The gravitational wave in this region is then specified by the functions $g$,
$V_{\scriptscriptstyle-}(g)$ and $M_{\scriptscriptstyle-}(g)$ which are again
related by a single equation corresponding to (\ref{Mf}), and by the single
parameter \ $n_{\scriptscriptstyle-}\ge2$ \ which is contained in the form of
$g(v)$. To set consistent initial conditions (and to ensure that
$M_{\scriptscriptstyle-}$ is at least $C^0$ across the wavefront), it is
necessary that the functions satisfy the condition 
 \begin{equation} 
 \lim_{\scriptstyle g\to 1/2}
\left[({\textstyle{1\over2}}-g) (V_{{\scriptscriptstyle-}g})^2\right]
=2k_{\scriptscriptstyle-}, 
 \label{shockcondition2}
 \end{equation}
 where \ ${1\over2}\le k_{\scriptscriptstyle-}<1$, \ (or \ 
$k_{\scriptscriptstyle-}=1-1/n_{\scriptscriptstyle-}$).

\medskip\goodbreak
It is always possible to choose initial $x$,$y$ coordinates such that the metric
on the wavefronts prior to the collision is continuous with the background metric
in the form (\ref{Minkowski}). It is therefore appropriate to impose the
additional initial conditions that \
$V_{\scriptscriptstyle+}({1\over2})=0$, \
$V_{\scriptscriptstyle-}({1\over2})=0$, \
$M_{\scriptscriptstyle+}({1\over2})=0$ \ and \
$M_{\scriptscriptstyle-}({1\over2})=0$.

{\bf Region~IV} ($u\ge0$, $v\ge0$):
In the vacuum case for the collision of plane gravitational waves with constant
aligned polarization, this region can always be described by the line element 
 \begin{equation}
 \d s^2=2e^{-M}\d u\,\d v-e^{-U}\big(e^{V}\,\d x^2 +e^{-V}\,\d y^2\big)
 \label{metric}
 \end{equation}
 where $U$, $V$ and $M$ are functions of both $u$ and $v$. The field equations
immediately imply that $e^{-U}$ satisfies a wave equation, so that we can put
 \begin{equation}
 e^{-U}=f(u)+g(v), 
 \label{alpha}
 \end{equation}
 where $f(u)$ and $g(v)$ are arbitrary functions subject to $f+g>0$. However,
since $e^{-U}$ must be continuous across the boundaries $u=0$ and $v=0$, $f(u)$
and $g(v)$ must retain the same forms as have already been prescribed in the
initial regions. If the special gauge (\ref{f}) and (\ref{g}) is adopted, it is
possible to make the global definitions 
 \begin{equation}
 f(u)={\textstyle{1\over2}}
-(c_{\scriptscriptstyle+}u)^{n_{\scriptscriptstyle+}}\Theta(u),
\qquad g(v)={\textstyle{1\over2}}
-(c_{\scriptscriptstyle-}v)^{n_{\scriptscriptstyle-}}\Theta(v) 
 \label{fg}
 \end{equation}
 where $\Theta(u)$ is the Heaviside step function, but this prescription is not
essential (and is often not convenient).

It may immediately be observed that a singularity of some kind will occur in the
interaction region on the spacelike hypersurface on which \ $f+g=0$. \ It can
be shown (see references in \cite{Griff91}) that this is generically a curvature
singularity, although families of solutions with infinitely many parameters exist
in which it corresponds to an unstable Cauchy horizon. It is referred to as a
``focussing'' singularity in figure~1.

In complete generality, we can adopt $f$ and $g$ as coordinates throughout
the interaction region, in which \ $f<{1\over2}$, \ $g<{1\over2}$ \ and \
$f+g>0$. \ With this, the main vacuum field equation is
 \begin{equation}
(f+g)V_{fg}+{\textstyle{1\over2}}V_f+{\textstyle{1\over2}}V_g=0,
 \label{EPD}
 \end{equation}
 which is an Euler--Poisson--Darboux equation with non-integer coefficients. For
any solution of this equation, $M$ can be obtained, at least in principle, by
integrating the remaining field equations 
 \begin{eqnarray}
 M_f &=&-{f''\over{f'}^2} +{1\over2(f+g)} -{\textstyle{1\over2}}(f+g){V_f}^2
 \label{M1}\\
 M_g &=&-{g''\over{g'}^2} +{1\over2(f+g)} -{\textstyle{1\over2}}(f+g){V_g}^2
 \label{M2} 
 \end{eqnarray}

In this case, the initial data is defined by the functions $f$, $g$, 
$V_{\scriptscriptstyle+}(f)$ and $V_{\scriptscriptstyle-}(g)$ and the solution of
(\ref{EPD}) in the interaction region $V(f,g)$ must satisfy \
 \begin{eqnarray}
 V(f,{\textstyle{1\over2}})=V_{\scriptscriptstyle+}(f), 
\qquad &&V({\textstyle{1\over2}},g)=V_{\scriptscriptstyle-}(g). 
 \end{eqnarray}
 with \ $V({1\over2},{1\over2})=0$. \ If these conditions are satisfied, the
required continuity of $M$ is assured.  With given initial data, these conditions
are sufficient to uniquely determine the solution of the colliding plane wave
problem in the interaction region, at least up to the focussing singularity.  We
recall that, for colliding plane waves, the metric functions must be continuous
across the wavefronts, while $V_f$ must be unbounded on \ $f={1\over2}$ \ but
constrained by (\ref{shockcondition}). Similarly, $V_g$ must be unbounded on \
$g={1\over2}$ \ subject to (\ref{shockcondition2}). However, these conditions are
already included in the initial data functions $V_{\scriptscriptstyle+}(f)$ and
$V_{\scriptscriptstyle-}(g)$.

\section{The Abel transform solution to the initial value problem}

A separable solution of equation (\ref{EPD}) can easily be obtained in the form
 \begin{equation}
 V(f,g)=A(\sigma-f)^{-{1\over2}}(\sigma+g)^{-{1\over2}} 
 \end{equation}
 where $A$ and $\sigma$ are arbitrary constants. Since many such functions can be
superposed, we can replace $A$ by \
$A(\sigma)\sqrt{\sigma+{\scriptstyle{1\over2}}}$ \ and treat $\sigma$ as a
``spectral'' parameter and $A(\sigma)$ as a spectral amplitude function.
Introducing also a second function $B(\sigma)$, a general class of solutions for
the interaction region~IV can be expressed in the form 
 \begin{equation}
 V(f,g)=\int_f^{1\over2} {A(\sigma)\sqrt{\sigma+{\scriptstyle{1\over2}}}
\over\sqrt{\sigma-f}\sqrt{\sigma+g}}\,\d\sigma
 +\int_g^{1\over2} {B(\sigma)\sqrt{\sigma+{\scriptstyle{1\over2}}}
\over\sqrt{\sigma+f}\sqrt{\sigma-g}}\,\d\sigma. 
 \label{gensolution}
 \end{equation}
 By considering the limits of this arbitrarily close to the two wavefronts as \
$g\to{1\over2}$ \ and \ $f\to{1\over2}$, \ it can be seen that this solution is
consistent with the initial data provided 
 \begin{equation}
 V_{\scriptscriptstyle+}(f) =
 \int_f^{1\over2} {A(\sigma)\over\sqrt{\sigma-f}}\,\d\sigma, \qquad
 V_{\scriptscriptstyle-}(g) =
\int_g^{1\over2} {B(\sigma)\over\sqrt{\sigma-g}}\,\d\sigma. 
 \label{Abel1}
 \end{equation}  
 As observed by Hauser and Ernst \cite{HauErn89a}, these integrals are exactly
those that define the Abel transform. It then follows that the spectral amplitude
functions $A(\sigma)$ and $B(\sigma)$ can be determined explicitly from the
initial data functions $V_{\scriptscriptstyle+}(f)$ and
$V_{\scriptscriptstyle-}(g)$ using the inverse Abel transform 
 \begin{equation}
 A(\sigma)=-{1\over\pi}\int_\sigma^{1\over2}
{V_{\scriptscriptstyle+}'(f)\over\sqrt{f-\sigma}}\,\d f, \qquad
 B(\sigma)=-{1\over\pi}\int_\sigma^{1\over2}
{V_{\scriptscriptstyle-}'(g)\over\sqrt{g-\sigma}}\,\d g. 
 \label{Abel2}
 \end{equation} 
 When comparing (\ref{Abel1}) with (\ref{gensolution}), it may be observed that
the solution in the interaction region involves the introduction of a dependence
upon $g$ in the integral involving $A(\sigma)$ and a dependence upon $f$ in the
integral involving $B(\sigma)$. This is an initial indication of a kind of
``dynamical monodromy data'' (as will be required for the nonlinear case and
explained in \cite{AleGri01} and \cite{AleGri02}) in which initial spectral data
on one characteristic is modified by the presence of that on the other.

Given initial data in the form of the specified functions $f$, $g$,
$V_{\scriptscriptstyle+}(f)$ and $V_{\scriptscriptstyle-}(g)$, the derivatives
$V_{\scriptscriptstyle+}'(f)$  and $V_{\scriptscriptstyle-}'(g)$ can be computed
and substituted into (\ref{Abel2}). The spectral functions $A(\sigma)$ and
$B(\sigma)$ are then uniquely determined, and the solution throughout the
interaction region is given by~(\ref{gensolution}). This solves the initial value
problem for the linear case, at least theoretically. In practice, it depends on
whether or not the integrals for $A(\sigma)$ and $B(\sigma)$ in (\ref{Abel2}) and
$V(f,g)$ in (\ref{gensolution}) can be explicitly evaluated.

It is possible that $V_{\scriptscriptstyle+}(f)$ and $V_{\scriptscriptstyle-}(g)$
may be expressed as a sum of distinct terms. It may then be observed that the
Abel transform method (\ref{gensolution})--(\ref{Abel2}) is linear. Thus, each
component of $V_{\scriptscriptstyle\pm}$ will give rise to separate components of
the spectral functions $A(\sigma)$ and $B(\sigma)$. These different components
will give rise to different components of the solution in the interaction region.
Since the main field equation (\ref{EPD}) is also linear, the complete solution
may be obtained as a simple sum over all these components.

Of course it is not necessary to construct the solution in the interaction
region using a single integral. In cases in which the expressions change
over different intervals (e.g. for sandwich waves, or waves with successive
impulsive components), it is appropriate to divide the interaction region
into a number of sub-regions and to determine the solution in each region
successively. Initial data can be specified on any pair of null characteristics.
Starting from any particular wavefront given by $f=f_1$, the Abel transform given
above can immediately be generalised to 
 \begin{equation}
 V_{\scriptscriptstyle+}(f) =V_{\scriptscriptstyle+}(f_1)
+\int_f^{f_1} {A(\sigma)\over\sqrt{\sigma-f}}\,\d\sigma, \qquad
A(\sigma)=-{1\over\pi}\int_\sigma^{f_1}
{V_{\scriptscriptstyle+}'(f)\over\sqrt{f-\sigma}}\,\d f. 
 \end{equation} 
 This enables the Abel transform method to be used in a sequence of
regions.

In particular, the solution of the characteristic initial value problem, in
which data are given on the two characteristics \ $f=f_1$\ and \ $g=g_1$, \
is given by first determining the spectral data 
 \begin{equation}
 A(\sigma)=-{1\over\pi}\int_\sigma^{f_1}
{V_{\scriptscriptstyle+}'(f,g_1)\over\sqrt{f-\sigma}}\,\d f, \qquad 
 B(\sigma)=-{1\over\pi}\int_\sigma^{g_1}
{V_{\scriptscriptstyle-}'(f_1,g)\over\sqrt{g-\sigma}}\,\d g. 
 \end{equation}
 The solution can then be obtained by evaluating 
 \begin{equation}
 V(f,g)=V(f_1,g_1)
+\int_f^{f_1} {A(\sigma)\over\sqrt{\sigma-f}}
{\sqrt{\sigma+g_1}\over\sqrt{\sigma+g}}\,\d\sigma
 +\int_g^{g_1} {B(\sigma)\over\sqrt{\sigma-g}}
{\sqrt{\sigma+f_1}\over\sqrt{\sigma+f}} \,\d\sigma. 
 \end{equation}

\section{Illustrations of the Abel transform method}

In this section, we will illustrate how the above method can be used explicitly
to construct some specific solutions.

As a first example, we will derive the Khan--Penrose solution \cite{KhaPen71}
which describes the collision of impulsive gravitational waves with aligned
polarization. In this case, \ $h_{\scriptscriptstyle+}(u)=\delta(u)$ \ so that \
$P_{\scriptscriptstyle+}=1-u\Theta(u)$, \ $Q_{\scriptscriptstyle+}=1+u\Theta(u)$,
\ and the solution in region~II is given by 
 \begin{equation}
 e^{-U_+}=1-u^2, \qquad e^{V_+}={1-u\over1+u}. 
 \end{equation} 
 Thus $n_{\scriptscriptstyle+}=2$, and we may put 
 \begin{equation}
 f={\textstyle{1\over2}}-u^2, \qquad {\rm and} \qquad
V_{\scriptscriptstyle+}(f)=\log\left(1-\sqrt{{\textstyle{1\over2}}-f}\right)
-\log\left(1+\sqrt{{\textstyle{1\over2}}-f}\right). 
 \label{KP+}
 \end{equation}
 Similarly in region~III, $h_{\scriptscriptstyle-}(v)=\delta(v)$, and the solution
is given by 
 \begin{equation}
 e^{-U_-}=1-v^2, \qquad e^{V_-}={1-v\over1+v}, 
 \end{equation}
 so that $n_{\scriptscriptstyle-}=2$, and we can put 
 \begin{equation}
 g={\textstyle{1\over2}}-v^2, \qquad {\rm and} \qquad
V_{\scriptscriptstyle-}(g)=\log\left(1-\sqrt{{\textstyle{1\over2}}-g}\right)
-\log\left(1+\sqrt{{\textstyle{1\over2}}-g}\right). 
 \label{KP-}
 \end{equation}

We may now use the Abel transform to determine the ``spectral'' functions
$A(\sigma)$ and $B(\sigma)$ that are required in the above solution of the
(linear) initial value problem. In this case, using initial data on the null
characteristic $v=0$, $u\ge0$, we obtain 
 \begin{eqnarray}
 A(\sigma)&=& -{1\over\pi} \int_\sigma^{1\over2}
{1\over\sqrt{f-\sigma}\sqrt{{1\over2}-f}\,({1\over2}+f)}\,\d f \nonumber\\  
 &=& {1\over\pi} \left[ {2\over\sqrt{{1\over2}+\sigma}} \tan^{-1} \left(
{\textstyle\sqrt{{1\over2}+\sigma}} \>
{\sqrt{{1\over2}-f}\over\sqrt{f-\sigma}} \right)
\right]_\sigma^{1\over2} \nonumber\\ 
 &=& -\left({\textstyle{1\over2}}+\sigma\right)^{-1/2}. 
 \end{eqnarray}
 By the identical calculation, using the
initial data on the other null characteristic $u=0$, $v\ge0$, we obtain
similarly that $B(\sigma)=-({1\over2}+\sigma)^{-1/2}$.

The solution in the interaction region may finally be obtained by evaluating 
 \begin{eqnarray}
 V(f,g) &=& -\int_f^{1\over2} {1\over\sqrt{\sigma-f}\sqrt{\sigma+g}}\,\d\sigma
-\int_g^{1\over2} {1\over\sqrt{\sigma+f}\sqrt{\sigma-g}}\,\d\sigma \nonumber\\ 
 &=& -\left[ \log\left( 2\sigma-f+g+2\sqrt{\sigma-f}\sqrt{\sigma+g} \right)
\right]_f^{1\over2} \nonumber\\ 
 &&\qquad\qquad -\left[ \log\left( 2\sigma+f-g+2\sqrt{\sigma+f}\sqrt{\sigma-g}
\right) \right]_g^{1\over2} \nonumber\\  
 &=& \log\left({\sqrt{{1\over2}+g}-\sqrt{{1\over2}-f}
\over\sqrt{{1\over2}+g}+\sqrt{{1\over2}-f}}\right) 
+\log\left({\sqrt{{1\over2}+f}-\sqrt{{1\over2}-g}
\over\sqrt{{1\over2}+f}+\sqrt{{1\over2}-g}}\right) 
 \label{KPsolution}
 \end{eqnarray}
 This is exactly the Khan--Penrose solution. To obtain the complete solution,
it only remains to integrate the subsidiary equations (\ref{M1}) and (\ref{M2})
to obtain the required expression for~$M$.

The above example is a particularly important illustration of the method for
constructing explicit solutions for given initial data since it describes
the collision of impulsive gravitational waves. For such waves, the metric
possesses the lowest possible degree of continuity across the wavefronts. In this
sense, the above solution represents the ``worst case'', and it is important to
observe that the above Abel transform method works perfectly adequately.

The above solution can immediately be generalised to that of Szekeres
\cite{Szek72} in which $n_{\scriptscriptstyle\pm}\ge2$ simply by multiplying
$V_{\scriptscriptstyle+}(f)$ and $V_{\scriptscriptstyle-}(g)$, and hence
$A(\sigma)$ and $B(\sigma)$, by the constants $\sqrt{2k_{\scriptscriptstyle+}}$
and $\sqrt{2k_{\scriptscriptstyle-}}$ respectively. For this case the two terms in
(\ref{KPsolution}) are multiplied by $\sqrt{2k_{\scriptscriptstyle+}}$
and $\sqrt{2k_{\scriptscriptstyle-}}$ respectively. However, for this solution, the initial
profiles are given by 
 \begin{equation}
 h_{\scriptscriptstyle+}(u) =
 \left\{ \begin{array}{lcl}
 \hskip6pc \delta(u) &{\rm if} \ n_{\scriptscriptstyle+}=2 \\
 \noalign{\medskip}
{\displaystyle {\sqrt{n_{\scriptscriptstyle+}
(n_{\scriptscriptstyle+}-1)} (n_{\scriptscriptstyle+}-2) \over2\sqrt2}\,
{u^{(n_{\scriptscriptstyle+}/2)-2}
\over(1-u^{n_{\scriptscriptstyle+}})^{3-(2/n_{\scriptscriptstyle+})}}\, \Theta(u)}
 \qquad &{\rm if} \ n_{\scriptscriptstyle+}>2
\end{array} \right.
 \end{equation}
 with equivalent expressions for $h_{\scriptscriptstyle-}(v)$. These have the
unphysical property that, apart from the Khan--Penrose case, the wave amplitude
becomes unbounded in the initial regions as \ $u\to1$, \ or \ $f\to-{1\over2}$.

The above example is a member of a family of solutions of (\ref{EPD}) that can be
expressed as products of Legendre functions. For these it is convenient to
introduce new coordinates $t$ and $z$ defined by 
 \begin{eqnarray}
 t&=& \sqrt{{\textstyle{1\over 2}}-f} \sqrt{{\textstyle{1\over 2}}+g}
+\sqrt{{\textstyle{1\over 2}}-g} \sqrt{{\textstyle{1\over 2}}+f} \nonumber\\
 z&=& \sqrt{{\textstyle{1\over 2}}-f} \sqrt{{\textstyle{1\over 2}}+g}
-\sqrt{{\textstyle{1\over 2}}-g}
\sqrt{{\textstyle{1\over 2}}+f}. 
 \end{eqnarray}
 The general family of solutions is then expressed as the sum 
 \begin{equation}
 V=\sum_n\Big(a_nP_n(t)P_n(z)+q_nQ_n(t)P_n(z)+p_nP_n(t)Q_n(z)
+b_nQ_n(t)Q_n(z) \Big) 
 \label{LegendreExpansion}
 \end{equation} 
 where $P_n(x)$ and $Q_n(x)$ are Legendre functions of the first and second
kinds respectively, and $a_n$, $q_n$, $p_n$ and $b_n$ are series of
arbitrary constants. For example, the Szekeres solution is given by \
 $V=-\sqrt2(\sqrt k_{\scriptscriptstyle+} +\sqrt k_{\scriptscriptstyle-})
Q_0(t)P_0(z) -\sqrt2(\sqrt k_{\scriptscriptstyle+} -\sqrt k_{\scriptscriptstyle-})
P_0(t)Q_0(z)$, \
 which also includes the Khan--Penrose solution. To satisfy the colliding plane
wave conditions (\ref{shockcondition}) and (\ref{shockcondition2}), which require
that the derivative of $V$ with respect to $f$ and $g$ be unbounded on the
wavefront, it is necessary that at least one Legendre function of the second kind
be included.

When considering the initial data that gives rise to these solutions, the
associated spectral functions can readily be calculated, and the solution
corresponding to each component can be determined, at least in the simpler cases.
For example, the spectral functions for the initial data associated with the
following components are: 
 \begin{equation}
\begin{array}{lll}
 V=Q_0(t)P_0(z): \qquad\qquad
 &A(\sigma) ={\displaystyle{1\over2\sqrt{{1\over2}+\sigma}}}, \qquad\qquad
 &B(\sigma) ={\displaystyle{1\over2\sqrt{{1\over2}+\sigma}}}, \\ 
 \noalign{\medskip}
 V=Q_1(t)P_1(z): \qquad\qquad
 &A(\sigma) =-{\displaystyle{\sigma\over\sqrt{{1\over2}+\sigma}}}, \qquad\qquad
 &B(\sigma) ={\displaystyle{\sigma\over\sqrt{{1\over2}+\sigma}}},\\ 
 \noalign{\medskip}
 V=Q_2(t)P_2(z): \qquad\qquad
 &A(\sigma) =-{\displaystyle{12\sigma^2-1 \over4\sqrt{{1\over2}+\sigma}}},
\qquad\qquad
 &B(\sigma) =-{\displaystyle{12\sigma^2-1 \over4\sqrt{{1\over2}+\sigma}}},\\ 
 \noalign{\medskip}
 V=Q_3(t)P_3(z): \qquad\qquad
 &A(\sigma) ={\displaystyle{20\sigma^3-3\sigma \over2\sqrt{{1\over2}+\sigma}}},
\qquad\qquad
 &B(\sigma) =-{\displaystyle{20\sigma^3-3\sigma \over2\sqrt{{1\over2}+\sigma}}}.
\end{array} 
 \end{equation} 
 The complete solution corresponding to these functions can readily be evaluated.
However, it may be observed that, for the same reasons as those outlined above,
combinations of these solutions do not generally correspond to realistic initial
data.

Another disadvantage of the expansion (\ref{LegendreExpansion}) is that the $f$
and $g$ derivatives of all the components which involve Legendre functions of the
second kind are unbounded on the wavefronts. Thus, the structure of the wavefronts
are represented by an unlimited number of these terms. All the coefficients appear
in the conditions (\ref{shockcondition}) and (\ref{shockcondition2}), and so are
constrained by the two constants $n_\pm$.

\section{Some difficulties which arise in practice}

It may be noted that, for sandwich waves or sequences of impulsive waves,
$h_{\scriptscriptstyle+}(u)$ is nonzero over a short interval only. If we denote
the end of this interval by \ $u=u_1$, \ we can consider the case in which \
$h_{\scriptscriptstyle+}(u)=0$ \ for \ $u>u_1$. \ Over this subsequent interval,
(\ref{PQdef}) implies that $P_{\scriptscriptstyle+}(u)$ and
$Q_{\scriptscriptstyle+}(u)$ must be linear in $u$, and it can be shown that the
spectral function must take the form 
 \begin{equation}
 A(\sigma) = -{2\over\pi\sqrt{\sigma+{\scriptstyle{1\over2}}}} \tan^{-1}
\left( c\, {\sqrt{f_1-\sigma}\over \sqrt{\sigma+{\scriptstyle{1\over2}}}} 
\right), 
 \end{equation} 
 where \ $f_1=f(u_1)$, \ and $c$ is some constant. Although the initial data can
be reconstructed from this expression using (\ref{Abel1}), the metric in the
interaction region can only be evaluated explicitly using (\ref{gensolution}) for
the case of the Khan--Penrose solution for which \ $c=\infty$. \ For all other
cases, (\ref{gensolution}) gives rise to integrals of elliptic integrals of the
third kind. For such situations it is most unlikely that the exact solution in the
interaction region can be expressed using a finite number of expressions involving
only elementary functions.

Using (\ref{condition1}), it may be noted that the profile function (\ref{h(u)1})
can be expressed in the form  
 \begin{equation}
 h_{\scriptscriptstyle+}(u) = {\textstyle
-{1\over2}(V_{{\scriptscriptstyle+},uu}-U_{{\scriptscriptstyle+},uu})
-{1\over4}(V_{{\scriptscriptstyle+},u}-U_{{\scriptscriptstyle+},u})^2 }. 
 \label{h(u)2}
 \end{equation}
 From this, it would appear to be natural to include $U_{\scriptscriptstyle+}$ as
one component of $V_{\scriptscriptstyle+}$. However, the spectral function for
this component takes the form 
 \begin{equation}
 A(\sigma) ={2\over\pi} {1\over\sqrt{{1\over2}+\sigma}} \>
\tan^{-1}\sqrt{{{1\over2}-\sigma\over{1\over2}+\sigma}} .
 \end{equation}
 Although the initial data can be reconstructed from this expression,
(\ref{gensolution}) again gives rise to integrals of elliptic integrals of the
third kind which cannot be evaluated explicitly in terms of elementary functions.
(The degenerate Ferrari--Iba\~nez solution \cite{FerIba88}, for which the
interaction region is locally isometric to part of the Schwarzschild solution
inside the horizon, is exactly of this form. It corresponds to the case in which
$V_{\scriptscriptstyle\pm}$ is the sum of the Khan--Penrose expression
(\ref{KP+}) or (\ref{KP-}) and $U_{\scriptscriptstyle\pm}$.)

It may be noted that (\ref{h(u)1}) can also be expressed in the form 
 \begin{equation}
 h_{\scriptscriptstyle+}(u) = -{\textstyle{1\over2}} V_{{\scriptscriptstyle+},uu}
-{f'\,V_{{\scriptscriptstyle+},u}\over2(f+{1\over2})} 
 \label{h(u)3}
 \end{equation}
 from which it is clear that, for most initial choices of
$V_{\scriptscriptstyle+}(u)$, $h_{\scriptscriptstyle+}(u)$ must be unbounded as \
$f\to-{1\over2}$. \ This feature is indeed observed for almost all the ``known''
solutions for colliding plane waves (including the degenerate Ferrari--Iba\~nez
solution). For the initial value problem considered here, however, it is
considered preferable that this singularity in the initial data does not occur.
We prefer to assume that realistic expressions for $V_{\scriptscriptstyle+}(u)$
should normally take a form in which such a singularity is avoided. However,
apart from the Khan--Penrose solution, none of the known families of solutions of
(\ref{EPD}) for $V_{\scriptscriptstyle+}(u)$ which satisfy the condition
(\ref{shockcondition}) also satisfy this regularity condition.

These observations would seem to argue that, for collisions of plane waves with
reasonably realistic expressions for $h_{\scriptscriptstyle+}(u)$ and
$h_{\scriptscriptstyle-}(v)$, it is most unlikely that exact solutions
representing the interaction region will be obtainable in any explicit form
involving elementary functions.

\section{Series of self-similar solutions}

We may, however, recall that there exists a family of self-similar solutions of
(\ref{EPD}) that have been described in \cite{AleGri95}--\cite{AleGri97} which
represent waves with distinct wavefronts. These were originally presented in the
context of different backgrounds, and in a form in which the wavefront \ $u=0$ \
is given by \ $f=0$. \ We will now demonstrate how this approach can be applied to
the colliding plane wave problem.

In order to consider the wave with the wavefront \ $u=0$, \ $f={1\over2}$, \ we
replace the coordinates $f$ and $g$ in \cite{AleGri95}--\cite{AleGri97} by
$f-{1\over2}$ and $g+{1\over2}$ respectively. We then introduce new coordinates
$\tau,\zeta$ in the interaction region defined by 
 \begin{equation}
 \tau=f+g, \qquad \zeta={1-f+g\over f+g}, 
 \end{equation} 
 In terms of the gauge (\ref{fg}), these take the form 
 \begin{equation}
 \tau=1 -(c_{\scriptscriptstyle+}u)^{n_{\scriptscriptstyle+}}
-(c_{\scriptscriptstyle-}v)^{n_{\scriptscriptstyle-}}, \qquad
 \zeta={1 +(c_{\scriptscriptstyle+}u)^{n_{\scriptscriptstyle+}}
-(c_{\scriptscriptstyle-}v)^{n_{\scriptscriptstyle-}} \over 1
-(c_{\scriptscriptstyle+}u)^{n_{\scriptscriptstyle+}}
-(c_{\scriptscriptstyle-}v)^{n_{\scriptscriptstyle-}}}, 
 \end{equation} 
 although this restriction is not necessary. Using the new coordinates, the
wavefront is given by \ $\zeta=1$, \ and a solution of (\ref{EPD}) can be
expressed in the self-similar form 
 \begin{equation}
 V(\tau,\zeta)=\tau^k H_k(\zeta) 
 \end{equation}
 where $k$ is an arbitrary real (non-negative) parameter. The condition that
\ $V=0$ \ on this wavefront is expressed by the constraint \ $H_k(1)=0$ \
for \ $k\ge0$. \ By direct substitution, it is found that the functions
$H_k(\zeta)$ must satisfy the linear ordinary differential equation 
 \begin{equation}
 (1-\zeta^2) H_k''+(2k-1)\zeta H_k'-k^2 H_k = 0. 
 \end{equation} 
 With the above initial conditions, these functions satisfy the recursion
relations 
 \begin{equation}
 H_k(\zeta) =\int_1^\zeta H_{k-1}(\zeta')\,d\zeta' \qquad
{\rm so\ that} \qquad H_k'(\zeta)=H_{k-1}(\zeta). 
 \end{equation} 
 For integer values of $k$, explicit expressions for these solutions can be
obtained in terms of elementary functions from the initial solution \
$H_0(\zeta)=\cosh^{-1}\zeta$. \

The solutions described above can be expressed in terms of standard hypergeometric
functions $F\big(a,b\,;c\,;z\big)$ in the form 
 \begin{equation}
 (f+g)^k H_k\left({\textstyle {1-f+g\over f+g} }\right) 
= c_k\,{({1\over2}-f)^{1/2+k}\over\sqrt{f+g}}\,
F\left({\textstyle{1\over2}},{\textstyle{1\over2}}\,;
{\textstyle{3\over2}}+k\,;{f-{1\over2}\over{f+g}}\right) 
 \end{equation} 
 where, for integer $k$ 
 \begin{equation}
 c_k= (-1)^k {2^k\Gamma({\textstyle{3\over2}})
\over\Gamma(k+{\textstyle{3\over2}})}. 
 \end{equation} 
 When applying the above recurrence relation for arbitrary values of $k$, we only
require the recursion relation $c_{k-1}=-{1\over2}(k+{1\over2})c_k$.

It may be noted that the condition (\ref{shockcondition}) on the wavefront is
satisfied only for the case \ $k=0$, \ and then only with the additional
multiplicative constant \ $a_0=2\sqrt{2k_{\scriptscriptstyle+}}$. \ However, in
view of the linearity of this case, an arbitrary number of higher order terms may
also be included. In addition, the equivalent solutions with a distinct wavefront
which propagate in the opposite direction, and have the wavefront \ $v=0$ \ on
which \
$g={1\over2}$, \ can also be included. In this way, a general solution in the
interaction region can be expressed as 
 \begin{equation}
 V(f,g)= \sum_{k=0}^\infty a_k 
 (f+g)^k H_k\left({\textstyle {1-f+g\over f+g} }\right) 
+\sum_{k=0}^\infty b_k 
 (f+g)^k H_k\left({\textstyle {1+f-g\over f+g} }\right), 
 \label{gensolution2}
 \end{equation}
 where \ $a_0=2\sqrt{2k_{\scriptscriptstyle+}}$ \ and \
$b_0=2\sqrt{2k_{\scriptscriptstyle-}}$. \ It may be observed that these
series have the advantage that terms with higher values of $k$ have a higher
degree of differentiability on the wavefront. This may be contrasted with
other series representations (in terms of Legendre or Bessel functions
etc.) in which all terms affect the differentiability properties on the
wavefront. On the other hand, it may well not be possible to explicitly evaluate
the quadratures for the remaining metric function $M$. Nevertheless, many terms
can be evaluated using the methods described in \cite{AleGri95}--\cite{AleGri97}
and some qualitative properties may be deduced.

The solution described above may be considered to be valid throughout the
interaction region \ $f<{1\over2}$, \ $g<{1\over2}$ \ and \ $f+g>0$. \ It
can then be extended back to region~II, according to the Penrose
construction, simply by setting \ $g={1\over2}$. \ The resulting expression
must then coincide with that prescribed by the initial data. Thus, 
 \begin{equation}
 V(f,{\textstyle{1\over2}})= \sum_{k=0}^\infty a_k 
 ({\textstyle{1\over2}}+f)^k 
H_k\left({\textstyle{{3\over2}-f\over{1\over2}+f}}\right)
 =V_{\scriptscriptstyle+}(f). 
 \end{equation} 
 Similarly, on the junction with region~III, we require 
 \begin{equation}
 V({\textstyle{1\over2}},g)= \sum_{k=0}^\infty b_k 
 ({\textstyle{1\over2}}+g)^k 
H_k\left({\textstyle{{3\over2}-g\over{1\over2}+g}}\right)
 =V_{\scriptscriptstyle-}(g). 
 \end{equation} 
 The given initial data $V_{\scriptscriptstyle+}(f)$ and
$V_{\scriptscriptstyle-}(g)$ can therefore be re-expressed in terms of two
sequences of constants $a_k$ and $b_k$. However, to address the initial
value problem, we need to find a way of determining the two sets of
constants $a_k$ and $b_k$ from the initial data functions
$V_{\scriptscriptstyle+}(f)$ and $V_{\scriptscriptstyle-}(g)$ respectively.

Once the coefficients $a_k$ and $b_k$ are known, the spectral functions
associated with each component can be obtained by evaluating 
 \begin{equation}
 A_k(\sigma) =-{1\over\pi}\int_\sigma^{1\over2}
{V_{{\scriptscriptstyle+}k}'(f)\over\sqrt{f-\sigma}}\,\d f, 
\qquad {\rm where} \qquad
V_{{\scriptscriptstyle+}k}(f) =a_k ({\textstyle{1\over2}}+f)^k 
H_k\left({\textstyle{{3\over2}-f\over{1\over2}+f}}\right), 
 \label{Ak}
 \end{equation}
 and similar expressions for $B_k(\sigma)$ and $V_{{\scriptscriptstyle-}k}(g)$
involving the constant $b_k$.
 It can be seen, however, that the integral (\ref{Ak}) must take the form 
 \begin{equation}
 A_k(\sigma) =p_k ({\textstyle{1\over2}}-\sigma)^k
({\textstyle{1\over2}}+\sigma)^{-1/2}, 
 \end{equation} 
 where $p_k$ is a constant that is to be determined. This result has been
obtained by substituting into the inverse transformation to reconstruct the
original component. Moreover, there is no difficulty in including also an
explicit $g$-dependence as required in the general expression
(\ref{gensolution}). Explicitly, the component in the interaction
region corresponding to $A_k(\sigma)$ is 
 \begin{eqnarray}
 V_k(f,g) &=& p_k \int_f^{1\over2}
{\left({\textstyle{1\over2}-\sigma}\right)^k
\over\sqrt{\sigma-f}\sqrt{\sigma+g}}\,\d\sigma \nonumber\\
 &=& p_k { \left({\textstyle{1\over2}}-f\right)^{{1\over2}+k}
\over\sqrt{f+g}}
\int_0^1 t^{-1/2}(1-t)^k
 \left(1+\left({\textstyle{{1\over2}-f\over f+g}}\right)t\right)^{-1/2} 
\>\d t \nonumber\\
 &=& p_k 
{\Gamma({1\over2})\Gamma(1+k)\over\Gamma({3\over2}+k)} \,
{ \left({\textstyle{1\over2}}-f\right)^{{1\over2}+k}
\over\sqrt{f+g}} \,
F\left({\textstyle{1\over2}},{\textstyle{1\over2}}\,;
{\textstyle{3\over2}}+k\,;-{{1\over2}-f\over{f+g}}\right) 
 \end{eqnarray} 
 which is exactly of the required form with 
 \begin{equation}
 p_k= (-1)^k {2^{k-1}\over\Gamma(1+k)} \,a_k . 
 \label{pk}
 \end{equation} 
 The spectral functions for the complete solution of the colliding plane
wave problem expressed in terms of these functions are therefore given by 
 \begin{equation}
 A(\sigma) =\sum_{k=0}^\infty p_k ({\textstyle{1\over2}}-\sigma)^k
({\textstyle{1\over2}}+\sigma)^{-1/2}, \qquad
B(\sigma) =\sum_{k=0}^\infty q_k ({\textstyle{1\over2}}-\sigma)^k
({\textstyle{1\over2}}+\sigma)^{-1/2}, 
 \end{equation} 
 where \ $p_0=\sqrt{2k_{\scriptscriptstyle+}}$ \ and \
$q_0=\sqrt{2k_{\scriptscriptstyle-}}$.

Remarkably, this provides a feasible way to solve the above characteristic initial
value problem. It is not necessary to determine the two sets of constants $a_k$
and $b_k$ from the initial data functions $V_{\scriptscriptstyle+}(f)$ and
$V_{\scriptscriptstyle-}(g)$. Rather, it is only necessary to determine the
spectral functions from the initial data. The functions \
$A(\sigma)\sqrt{{1\over2}+\sigma}$ \ and \ $B(\sigma)\sqrt{{1\over2}+\sigma}$ \
should then be expanded as power series in \ ${1\over2}-\sigma$. \ The resulting
coefficients are exactly the coefficients $p_k$ and $q_k$ from which the
coefficients $a_k$ and $b_k$ can be determined using (\ref{pk}) and the
equivalent expression for $q_k$. In this way, the general solution
(\ref{gensolution2}) in the interaction region has been constructed explicitly,
albeit in terms of a pair of infinite series.

\section{Observations}

It has been demonstrated above that the initial value problem for colliding plane
waves has been solved in principle for the vacuum case in which the approaching
gravitational waves have constant and aligned polarization. The initial data are
specified by two functions $h_{\scriptscriptstyle+}(u)$ and
$h_{\scriptscriptstyle-}(v)$ which describe the profiles of the approaching
waves. The first step is to solve the differential equations (\ref{PQdef}) for
$P_{\scriptscriptstyle\pm}$ and $Q_{\scriptscriptstyle\pm}$. From these, the
functions $U_{\scriptscriptstyle+}(u)$ and $U_{\scriptscriptstyle-}(v)$ are
determined, and this specifies the functions $f(u)$ and $g(v)$. They also
determine the functions $V_{\scriptscriptstyle+}(u)$ and
$V_{\scriptscriptstyle-}(v)$, and the next step is to construct the spectral data
functions $A(\sigma)$ and $B(\sigma)$ using (\ref{Abel2}). From these, the final
solution for $V(f,g)$ in the interaction region is obtained by evaluating the
integrals in (\ref{gensolution}).

The method described above is correct in principle. In practice, however,
difficulties will occur at several stages. The greatest difficulties in the
construction of exact solutions will arise in the explicit evaluation of the
integrals for $A(\sigma)$ and $B(\sigma)$ in (\ref{Abel2}), and then particularly
in determining the final expression for $V(f,g)$ using (\ref{gensolution}). Also,
although expressions for $f(u)$ and $g(v)$ are determined explicitly using the
above procedure, it is not often that they can be inverted to construct
$V_{\scriptscriptstyle+}(f)$ and $V_{\scriptscriptstyle-}(g)$. Nevertheless, it
is possible to continue to use $u$ and $v$ as coordinates throughout the above
calculations.

Of course, the above procedure could be used as the basis for a numerical
solution of the problem. Indeed, for the collision of two pairs of impulsive
waves or two sandwich waves, this may be the only viable option. For such an
approach, care needs to be taken in evaluating the integrals, particularly near the
limits as the integrands there behave as $|f-\sigma|^{-1/2}$, and
$V'_{\scriptscriptstyle+}({1\over2})$ and $V'_{\scriptscriptstyle-}({1\over2})$
are also unbounded.

For the linear case considered above, the metric is diagonal and the solution is
derived with the aid of the Abel transform. Unfortunately, this approach cannot be
extended to the more general case in which the waves are not colinear. In the
general case, the approaching gravitational waves either have variable or
nonaligned polarization or electromagnetic wave components are included. In either
case the main field equations are essentially nonlinear. However, the field
equations are exactly the Ernst equations and their associated quadratures which
are known to be integrable. Nevertheless, none of the known solution-generating
techniques that are associated with these equations are adapted to initial data
on a pair of null characteristics. An alternative approach has therefore been
developed \cite{AleGri01}, \cite{AleGri02}. However, further work is required to
investigate whether or not this approach can deal in practice with realistic
initial data.

\section*{Acknowledgments}

For many useful discussions, the authors are particularly grateful to Professor
G. A. Alekseev, whose collaboration on this topic was partially supported by the
EPSRC and the INTAS grant 99-1782.

\end{document}